\documentstyle[amssymb,prl,aps,epsfig,twocolumn]{revtex}

\begin{document}
\draft

\twocolumn[\hsize\textwidth\columnwidth\hsize\csname@twocolumnfalse\endcsname
\title{Creating Steady Atom-Molecule Entanglement and Coherent Molecular output by Demkov-Kunike Type
 Non-adiabatic Transition }
\author{P. Zhang $^{1,2}$, H. T. Quan $^{1}$ and C. P. Sun $^{1,3,a,b}$}
\address{$^{1}$Institute of Theoretical Physics, the Chinese Academy of Sciences,
 Beijing, 100080, China\\
 $^{1}$ Department of Physics, the University of Hong Kong, Hong Kong, China\\
 $^{3}$Department of Physics, Nankai University, Tianjin, 300071, China}
\maketitle
\begin{abstract}
We, based on the photoassociation of fermion atoms into bosonic
molecules, propose a scheme to create the steady entanglement
between the atom state and the molecule state inside an optical
lattice. The stability of entanglement state is guaranteed by
sweeping the frequency of Ramman laser beam through resonance
according to the second Demkov-Kunike (DK2) nonadiabatic
transition model\cite{DK,models}. The probability amplitude of
each components can be precisely controlled by adjusting the
sweeping parameter. Considering the loss of molecule, the steady
coherent molecular output can also be obtained in each lattice
site.
\end{abstract}
 ]

In recent years, the photoassociation of fermionic atoms forming bosonic
molecules \cite{p6,p7,science,Mckenzie} has aroused great interest. In this
process, two atoms in condensate are illuminated by two Raman laser beams.
This atom pair absorbs an photon from one laser beam and emits a photon to
another beam. As a result of stimulated Raman effect, there is a transition
from the two atom state to a single molecule state. Most recently, there are
many experiments to demonstrate this photoassociation process and the
Bose-Einstein Condenstation (BEC) of molecular has been observed \cite%
{science}. There also exists another physical mechanism, Feshbach Resonance
to create the molecular from the very cold atoms even in situation of BEC.
When the external magnetic field of an atomic system is changed, an entrance
interaction channel and a closed channel can be resonantly coupled and the
population of a two atom bound state, i.e. the molecular sate, can be
resonantly enhanced. In recent experiments, the disappearance of atoms and
the creation of molecules from BEC have been observed \cite%
{nature1,nature2,rempe}.

In both the two cases, an interesting question is whether one can controll
exactly the rate of producing the composite particle--molecule and even
obtain the coherent output of molecular \cite{search}. If the dissipation
mechanism is not considered in this photassociation process, there must
exists a Rabi oscillation between atomic state and molecular state and then
there is only a "pulse profile" of molecular productions. In this case, we
can not obtain a steady population of atom and molecular state. This
phenomenon is very similar to "pulse atomic laser". In the experiment of
atomic laser, an crucial technology to overcome this problem is to sweep the
frequency of radiation field (r.f) \cite{atom laser} that couples the
trapped BEC state to the un-trapped state. This physical mechanism avoiding
the pure pulse profile due to Rabi oscillation can be understood in the
rotation picture. Within this representation, the sweeping frequency will
shift the original energy level of atom, Then the sweeping derives the
system to cross the degenerate point of energy level in a non-adiabatic way.
Then, the Landau-Zener approximation \cite{lz} can be utilized to obtain the
steady populations of BEC atoms in the different energy levels. In this
article, we will invoke this idea of sweeping frequency to obtain a steady
atom-molecule entanglement state, i.e. the coherence superposition of atomic
and molecular number states with steady population of molecules produced
from the atoms in photoassociation.

We notice that the Landau-Zener model for sweeping the frequency is
precisely correct, only when the detunning in the effective two level Rabi
model linearly varies with time $t$ from $-\infty $ to $+\infty $. Actually,
the linear function of time is only an approximate description of the
variation of the realistic detunning, because in practical cases the
experimental parameters are always adjusted as smooth functions of time
which have finite values in the limit $t\rightarrow \pm \infty $. For
instance, in ref. \cite{Kleppner}, the hyperbolic tangent function is
considered as a more practical description of the experimental behavior of
the atomic Stark energy difference controlled by laser beams. The validity
of the Landau Zener model is also analyzed numerically in the same reference.

In this article, we adopt a feasible sweeping way depicted by the second
Demkov-Kunike (DK2) \cite{DK,models} model. In this model, the detuning has
finite asymptotic values in the limit $t\rightarrow \pm \infty $. When we
sweep the detuning in the DK2 way, the finite asymptotic values of detuning
are experimentally implementable.

In our study, the model Hamiltonian for the atom-molecule photoassociation
inside a single site in a optical lattice can be written as \cite{weiping}:%
\begin{eqnarray}
H &=&\Omega _{b}b^{\dagger }b+\frac{1}{2}U_{b}b^{\dagger }b\left[ b^{\dagger
}b-1\right]  \label{a} \\
&&+\Omega _{f}\sigma _{z}+U_{x}b^{\dagger }b\sigma _{z}+\chi b^{\dagger
}\sigma _{-}+\chi b\sigma _{+}.  \nonumber
\end{eqnarray}%
Here, $b$ is the annihilation operator of molecule and $\chi $ is
proportional to the two-photon Rabi frequency associated with the two laser
beams \cite{weiping}. Without loss of generality, in the following
discussions, we assume $\chi $ to be real. $\Omega _{b}$ and $\Omega _{f}$
are defined as 
\begin{eqnarray*}
\Omega _{b} &=&\omega _{b}+\Delta +U_{x}, \\
\Omega _{f} &=&\omega _{f}+U_{f}/2
\end{eqnarray*}%
where $\omega _{b}$ is the atomic energy, $\omega _{f}$ is the molecular
energy and $\Delta $ the detunning of the two laser beams. The terms
contains $U_{f},$ $U_{x}$ and $U_{b}$ denote the energies of two body
interactions between atoms, atoms and molecules, and molecules. In the
Hamiltonian (\ref{a}), the Anderson's qusi-spin maps \cite{map} defined by 
\begin{equation}
\sigma _{-}=c_{1}c_{2},\sigma _{+}=c_{1}^{+}c_{2}^{+},\sigma
_{z}=c_{1}^{\dagger }c_{1}+c_{2}^{\dagger }c_{2}-1,
\end{equation}%
have been used. Here, $c_{i}$ $\left( i=1,2\right) $ is the annihilation
operator of the atomic state $\left\vert i\right\rangle $. It is easy to see
that the operators $\sigma _{\pm }$ are defined as the qusi-spin flips $%
\sigma _{+}=\left\vert e\right\rangle \left\langle g\right\vert $, $\sigma
_{-}=\left\vert g\right\rangle \left\langle e\right\vert $ and $\sigma
_{z}=\left\vert e\right\rangle \left\langle e\right\vert -\left\vert
g\right\rangle \left\langle g\right\vert $ where $\left\vert g\right\rangle
=\left\vert 0\right\rangle _{a}$ is the atom vacuum state and $\left\vert
e\right\rangle =c_{1}^{+}c_{2}^{+}\left\vert 0\right\rangle _{a}$ is the two
atom state. In the mathematical formulism, the Hamiltonian (\ref{a}) is just
a Jaynes-Cummings kind Hamiltonian with a nonlinear detuning \cite{weiping}.
The laser enhanced combination process depicted by the interaction
Hamiltonian can be illustrated in Fig 1.

\begin{figure}[h]
\caption{ The photoassociation process of the atoms in optical lattice. In
every lattice, the molecule state $|M\rangle $ (the two atom bound state) is
coupled with the atom state $|F\rangle $ (the two atom free state) via the
two Raman beams $\protect\omega _{1}$ and $\protect\omega _{2}$. }
\end{figure}

In order to use the DK2 model, we assume that the detunning $\Delta $ of the
two Raman laser beams varies with time as a hyperbolic tangent function: 
\begin{equation}
\Delta \left( t\right) =2\omega _{f}-\omega _{b}+U_{f}-2k\tanh \left( \frac{t%
}{T}\right) .  \label{b}
\end{equation}%
It is obviously that the asymptotic values of $\Delta \left( t\right) $ in
the limit $t=\pm \infty $ are determined by the parameter $k$ and the speed
of the variation of $\Delta \left( t\right) $ depends on $T$. Therefore, in
the limit $t\rightarrow \pm \infty $, the $\Delta $ has finite asympotic
values $2\omega _{f}-\omega _{b}+U_{f}\pm 2k$. Since $\left\vert \tanh
\left( \pm 2\right) \right\vert =0.96\approx 1$, the practical operational
time is about $4T$.

Obviously, the subspace expanded by $\left\{ \left\vert e,n_{b}\right\rangle
,\left\vert g,n_{b}+1\right\rangle \right\} $ is an invariance subspace $%
V_{n}$ of $H$ in Eq. (\ref{a}). In other words, the Hamiltonian can be
diagonalized as the block-diagonal matrix and an arbitrary superposition of $%
\left\vert e,n_{b}\right\rangle $ and $\left\vert g,n_{b}+1\right\rangle $
in $V_{n}$ can only evolve into $V_{n}$ itself. Here,\ we define $\left\vert
e,n_{b}\right\rangle =\left\vert e\right\rangle \otimes \left( n_{b}!\right)
^{-\frac{1}{2}}b^{\dagger }{}^{n_{b}}\left\vert 0\right\rangle _{b}$ as the
number state corresponding to two atoms and $n_{b}$ molecules. Here, $%
\left\vert 0\right\rangle _{b}$ is the molecule vacuum state. $\left\vert
g,n_{b}+1\right\rangle $ has the similar definition. Then a block of $H$ can
be expressed as a $2\times 2$ matrix:%
\begin{eqnarray}
H\left( n_{b},t\right)  &=&\left[ a+k\tanh \left( \frac{t}{T}\right) \right]
\left\vert e,n_{b}\right\rangle \left\langle e,n_{b}\right\vert +c\left\vert
e,n_{b}\right\rangle \left\langle g,n_{b}+1\right\vert   \label{i1} \\
&&-\left[ a+k\tanh \left( \frac{t}{T}\right) \right] \left\vert
g,n_{b}+1\right\rangle \left\langle g,n_{b}+1\right\vert +c\left\vert
g,n_{b}+1\right\rangle \left\langle e,n_{b}\right\vert   \nonumber
\end{eqnarray}

where we have used Eq. (\ref{b}). Here, $a$ and $c$ are functions of $n_{b}$
and are defined as $a\left( n_{b}\right) =n_{b}\left( U_{x}-U_{b}/2\right) ,$
$c\left( n_{b}\right) =\chi \sqrt{n_{b}+1}$.

Since $\chi $ is proportional to the Rabi frequency which is slowly varying
with the frequency of the Raman beams, it is reasonable to assum$%
9.2847701\times 10^{-24}%
\mathop{\rm J}%
\mathop{\rm T}%
^{-1}$e $c\left( n_{b}\right) =\chi \sqrt{n_{b}+1}$ to be a time-independent
constant. The instantaneous eigenvectors of the matrix $H\left(
n_{b},t\right) $ can be noted as $\left\vert \Phi _{+}\left( n_{b},t\right)
\right\rangle $ and $\left\vert \Phi _{-}\left( n_{b},t\right) \right\rangle 
$. It is apparently that in the limit $t\rightarrow \pm \infty $, the matrix 
$H\left( n_{b},t\right) $ has asymptotic expressions%
\begin{eqnarray*}
H\left( n_{b},\pm \infty \right)  &=&\left[ a\left( n_{b}\right) \pm k\right]
\left\vert e,n_{b}\right\rangle \left\langle e,n_{b}\right\vert +c\left(
n_{b}\right) \left\vert e,n_{b}\right\rangle \left\langle
g,n_{b}+1\right\vert  \\
&&-\left[ a\left( n_{b}\right) \pm k\right] \left\vert
g,n_{b}+1\right\rangle \left\langle g,n_{b}+1\right\vert +c\left(
n_{b}\right) \left\vert g,n_{b}+1\right\rangle \left\langle
e,n_{b}\right\vert .
\end{eqnarray*}%
Therefore, the eigenvectors $\left\vert \Phi _{\pm }\left( n_{b},t\right)
\right\rangle $ also have asymptotic expressions when $t\rightarrow \pm
\infty $. We note them as $\left\vert \Phi _{\pm }\left( n_{b},\pm \infty
\right) \right\rangle $. Generally speaking, $\left\vert \Phi _{\pm }\left(
n_{b},\pm \infty \right) \right\rangle $ are superposition states of $%
\left\vert e,n_{b}\right\rangle $ and $\left\vert g,n_{b}+1\right\rangle $.
It can be seen in the following discussion that we are only interested in
the special cases that $\left\vert \Phi _{\pm }\left( n_{b},\pm \infty
\right) \right\rangle $ can be approximated as $\left\vert
e,n_{b}\right\rangle $ or $\left\vert g,n_{b}+1\right\rangle $. Then the
general expression of $\left\vert \Phi _{\pm }\left( n_{b},t\right)
\right\rangle $ is not needed here.

The instantaneous state $\left\vert \Psi \left( n_{b},t\right) \right\rangle 
$ of the quantum system is the solution of the Schoredinger equation 
\[
i\frac{d}{dt}\left\vert \Psi \left( n_{b},t\right) \right\rangle =H\left(
n_{b},t\right) \left\vert \Psi \left( n_{b},t\right) \right\rangle 
\]%
and it can be expanded as the superposition of $\left\vert \Phi _{\pm
}\left( n_{b},t\right) \right\rangle $ 
\[
\left\vert \Psi \left( n_{b},t\right) \right\rangle =\sum_{i=+,-}D_{i}\left(
n_{b},t\right) \left\vert \Phi _{i}\left( n_{b},t\right) \right\rangle . 
\]%
Obviously, $\left\vert D_{\pm }\left( n_{b},t\right) \right\vert ^{2}$ are
the populations in the states $\left\vert \Phi _{\pm }\left( n_{b},t\right)
\right\rangle $ at any instantaneousness. In practical cases, the asymptotic
value $\left\vert D_{\pm }\left( n_{b},+\infty \right) \right\vert ^{2}$ in
the limit $t\rightarrow \pm \infty $ is usually of interest. If the quantum
system is initially prepared in the state%
\[
\left\vert \Psi \left( n_{b},-\infty \right) \right\rangle =\left\vert \Phi
_{+}\left( n_{b},-\infty \right) \right\rangle , 
\]%
the values of $\left\vert D_{\pm }\left( +\infty \right) \right\vert ^{2}$
can be obtained with direct calculations \cite{DK,models}:%
\begin{eqnarray}
\left\vert D_{-}\left( +\infty \right) \right\vert ^{2} &=&\frac{\sinh \left[
\pi TE_{+}\right] \sinh \left[ \pi TE_{-}\right] }{\sinh \left[ \pi TE_{a}%
\right] \sinh \left[ \pi TE_{e}\right] },  \label{e1} \\
\left\vert D_{+}\left( +\infty \right) \right\vert ^{2} &=&1-\left\vert
D_{-}\left( +\infty \right) \right\vert ^{2}  \nonumber
\end{eqnarray}%
where the complicated notations $E_{a}$, $E_{e}$ and $E_{\pm }$ are defined
as functions of the physical parameters $a$, $b$ and $k:$%
\begin{eqnarray}
E_{a}\left( a,k,c\right) &=&\left\vert \left[ \left( a-k\right) ^{2}+c^{2}%
\right] ^{\frac{1}{2}}\right\vert ,  \label{k1} \\
E_{e}\left( a,k,c\right) &=&\left\vert \left[ \left( a+k\right) ^{2}+c^{2}%
\right] ^{\frac{1}{2}}\right\vert ,  \nonumber \\
E_{\pm }\left( a,k,c\right) &=&k\pm \frac{E_{e}-E_{a}}{2}.  \nonumber
\end{eqnarray}%
In the above calculations, we have refered to the main result in the DK2
model.

To create the atom-molecule entanglement state from the direct product of
the atomic and molecular number sate $\left\vert e,n_{b}\right\rangle $, we
assume the amplitude $k$ is large enough so that the conditions $k\pm
a\left( n_{b}\right) >>c\left( n_{b}\right) $ are satisfied. Since $a$ is
proportional to $n_{b}$ and $c$ is proportional to $\sqrt{n_{b}}$, this
condition implies that $n_{b}$ is small enough. In this case, when $t=\pm
\infty $, since $k\pm a\left( n_{b}\right) >>c\left( n_{b}\right) $, the
diagonal terms of $H\left( n_{b},\pm \infty \right) $ is much larger than
the off diagonal terms. The eigenstates of $H\left( n_{b},\pm \infty \right) 
$ can be approximated as%
\begin{eqnarray*}
\left\vert \Phi _{+}\left( n_{b},-\infty \right) \right\rangle &\simeq
&\left\vert \Phi _{-}\left( n_{b},+\infty \right) \right\rangle \simeq
\left\vert e,n_{b}\right\rangle \\
\left\vert \Phi _{-}\left( n_{b},-\infty \right) \right\rangle &\simeq
&\left\vert \Phi _{+}\left( n_{b},+\infty \right) \right\rangle \simeq
\left\vert g,n_{b}+1\right\rangle
\end{eqnarray*}%
This is an analogy of large detunning case when a two level atom interact
with a classical radiation field. Therefore, in the limit $t\rightarrow \pm
\infty $, there is almost no atom-molecule transition and the mode of the
probability amplitudes of the states $\left\vert e,n_{b}\right\rangle $ and $%
\left\vert g,n_{b}+1\right\rangle $ do not change with time. This phenomenon
has been shown in the experiment of photoassociation \cite{science}. The
mixing of the two states mainly happens in the nearby of the "resonance
point" where $\left\vert a+k\tanh \left( t/T\right) \right\vert \lesssim c$.

In this case, if the system is initially prepared in the state $\left\vert
e,n_{b}\right\rangle $ when $t=-\infty $ and $\Delta $ is changed with time
according to condition (\ref{b}), the system will evolve into the
superposition state 
\begin{equation}
D_{1}\left( n_{b}\right) \left\vert e,n_{b}\right\rangle +D_{2}\left(
n_{b}\right) \left\vert g,n_{b}+1\right\rangle  \label{o1}
\end{equation}%
when $t=+\infty $. Apparently, this quantum state can be considered as a
atom-molecule entanglement state. It follows from Eq. (\ref{e1}) that%
\begin{equation}
\left\vert D_{1}\left( n_{b}\right) \right\vert ^{2}\approx \frac{\sinh %
\left[ \pi TE_{+}\left( a,k,c\right) \right] \sinh \left[ \pi TE_{-}\left(
a,k,c\right) \right] }{\sinh \left[ \pi TE_{a}\left( a,k,c\right) \right]
\sinh \left[ \pi TE_{e}\left( a,k,c\right) \right] }  \label{p1}
\end{equation}%
where the functions $E_{a}\left( a,k,c\right) $, $E_{e}\left( a,k,c\right) $
and $E_{\pm }\left( a,k,c\right) $ are defined in Eq. (\ref{k1}). The
probability of creating one molecule from two atoms can be expressed as $%
P=1-\left\vert D_{1}\right\vert ^{2}$. Especially when $n_{b}=0$, \thinspace 
$P$ describes the molecular vacuum production.

It is easy to prove that when $k$ is large enough so that $e^{Tk}>>1$, $%
\left\vert D_{1}\left( n_{b}\right) \right\vert ^{2}$ in Eq. (\ref{p1}) can
be approximated as%
\begin{eqnarray}
\left\vert D_{1}\left( n_{b}\right) \right\vert ^{2} &\approx &\exp \left[
-\pi \frac{Tc^{2}}{k\left( 1-\frac{a^{2}}{k^{2}}\right) }\right]  \label{q1}
\\
&=&\exp \left[ -2\pi \frac{c^{2}}{\frac{d}{dt}\left[ 2k\tanh \frac{t}{T}%
\right] |_{t=t_{0}}}\right] .  \nonumber
\end{eqnarray}%
Here, $t_{0}$ which satisfies $a+k\tanh \left( t_{0}/T\right) =0$ is the
cross point of the diagonal elements of the Hamiltonian in Eq. (\ref{i1}).
Therefore, the approximated expression of $\left\vert D_{1}\left(
n_{b}\right) \right\vert ^{2}$ in Eq. (\ref{q1}) is just the result given by
Landau Zener formula.

Now we are in a position to analyse the influence of the parameters $T$, $%
n_{b}$, $k$, $U_{x}$ and $U_{b}$ on the molecular probability $P$. It is
convenient to express the probability $P$ as $P=1-fg$ where $f=\frac{\sinh %
\left[ \pi TE_{-}\right] }{\sinh \left[ \pi TE_{a}\right] }$ and $g=\frac{%
\sinh \left[ \pi TE_{+}\right] }{\sinh \left[ \pi TE_{e}\right] }.$ Using
the simple relationships 
\begin{equation}
E_{a}-E_{-}=E_{e}-E_{+}=\frac{E_{e}+E_{a}}{2}-k>0,  \label{t1}
\end{equation}%
we can prove $df/dT<0$ and $dg/dT<0$ with the inequality $y\tanh x<x\tanh y$
when $x>y>0$. Then we have $dP/dT>0.$With straightforward calculations, it
is also proved that under the condition $k\pm a\left( n_{b}\right) >>c\left(
n_{b}\right) $, we have $dP/dn_{b}>0$ \cite{prove}. Therefore, the molecular
probability $P$ increases with $T$ and the molecular number $n_{b}$.
Therefore, in order to obtain a high molecular probability $P$, the speed of
the variation of the detuning $\Delta $ should be low enough. In Fig. 2, the
behavior of $P$ as a function of $T$ is shown. It can be seen from this
figure that when $n_{b}=0$ and $T$ is larger than $25\chi ^{-1}$, $P$ is
approximately unit and a molecule is created with unit probability from two
atoms. However, when $n_{b}=5$, we have $P\approx 1$ when $T$ is larger than 
$5\chi ^{-1}.$

\begin{figure}[h]
\caption{ The molecular probability $P$ as a function of $T$ in case the
molecular number $n_{b}=0,$ $2$ and $5.$ Here, we have assumed $k=20\protect%
\chi $, $\left( U_{x}-U_{b}/2\right) =0.5\protect\chi $ and the unit of $T$
is $\protect\chi ^{-1}$. It is obviously that $P$ increases with $T$ and $%
n_{b}$. }
\end{figure}

\bigskip The influences of the amplitude $k$ of the detunning and the atomic
scattering coefficient $\left( U_{x}-U_{b}/2\right) $ on the molecular
probability $P$ is shown in Fig. 3. It can be seen that with the same
operation time $T$ and molecular number $n_{b}$, as the amplitude $k$
increases, the molecular probability $P$ decreases. On the other hand, when
the scattering strength is small enough so that $\left( U_{x}-U_{b}/2\right)
/\chi $ $\lesssim 10$, $P$ has almost the same value with respect to
different values of $\left( U_{x}-U_{b}/2\right) $, i.e. the effect of
atomic scattering can be ignored. When $\left( U_{x}-U_{b}/2\right) $
becomes larger, $P$ increases with $\left( U_{x}-U_{b}/2\right) $. It is
pointed out that in Fig. 3 we assume $k\sim 10^{2}$ and $T\sim 10^{-4}$ so
that $e^{kT}<1$. Therefore, the figure is drawned in the domain where the
Landau-Zener formula is not applicable. In case of larger $k$ or $T$, the
same conclusion can be obtained with the expression of $P$ in Eq. (\ref{q1})
given by Landau-Zener formula.

\begin{figure}[h]
\caption{ The molecular probability $P$ as a function of $k$ with different
value of $\left( U_{x}-U_{b}/2\right) $. In the figure, we note $m=$ $\left(
U_{x}-U_{b}/2\right) .$ We assume $n_{b}=1$ and $T=10^{-2}\protect\chi ^{-1}$%
. The unit of $m$ and $k$ are all $\protect\chi $. The unit of $P$ is $%
10^{-3}$. }
\end{figure}

Apparently, the quantum state in Eq. (\ref{o1}) can also be considered as a
atom-molecule entanglement state written in the Schmit decomposition form.
The entanglement of this state can be described by the entropy of
entanglement $S=-\sum_{i=1,2}\left\vert D_{i}\left( n_{b}\right) \right\vert
\ln \left\vert D_{i}\left( n_{b}\right) \right\vert $ as a function of the
Schmit numbers $\left\vert D_{1}\left( n_{b}\right) \right\vert $ and $%
\left\vert D_{2}\left( n_{b}\right) \right\vert $. The variation of $S$ as a
function of $T$ has been shown in Fig. 4. It can be seen that the entropy $S$
has a non-zero value when $T=0$. This is because in the limit $T=0$, we have 
$P=1-E_{+}E_{-}/\left( E_{a}E_{e}\right) \neq 0.$ As $T$ increases, $S$
increases to its maximal value $0.49.$ In this case, the finial state is
just the maximal atom-molecule entanglement sate%
\begin{equation}
\left\vert \text{Bell}\right\rangle =\frac{1}{\sqrt{2}}\left[ \left\vert
e,n_{b}\right\rangle +\left\vert g,n_{b}+1\right\rangle \right] .  \label{u1}
\end{equation}%
After crossing its maximal value, the entropy $S$ slowly decreases to its
asymptotic value $0$ in the limit $T\rightarrow \infty $.

\begin{figure}[h]
\caption{ The entropy $S$ of entanglement as a function of $T$ with
different $n_{b}$. Here, we assume $k=10\protect\chi $ and $\left(
U_{x}-U_{b}/2\right) =0.5\protect\chi $. The unit of $T$ is $\protect\chi %
^{-1}$. }
\end{figure}

Now we are in a position to cosider the realization of steady coherent
molecular output with the nonadiabatic transition method. To this end we
assume a plused fermion beam with pluse lenght $T$ and period $\tau $
between each two pluses is used in our system. We assume in each inerval $%
\tau $ the two Raman laser beams is turnned on and the detuning $\Delta $ is
changed with time as in Eq. (\ref{b}). Therefore, if the state of the total
system is just $\left\vert e,n_{b}\right\rangle $ before the Raman beams are
turnned on, the transition from state $\left\vert e,n_{b}\right\rangle $ to
the state in Eq. (\ref{o1}) will happen during the time interval $\tau $. In
this case, if the loss of molecule is considered, it is known that the
evolution of the molecular population of the molecular field can be
described by the master equation \cite{weiping}:%
\begin{eqnarray}
\frac{d}{dt}p_{n} &=&-r_{a}\left[ 1-\left\vert D_{2}\left( n\right)
\right\vert ^{2}\right] p_{n}+r_{a}\left\vert D_{2}\left( n-1\right)
\right\vert ^{2}p_{n-1}  \label{u2} \\
&&-\gamma np_{n}+\gamma \left( n+1\right) p_{n+1}.  \nonumber
\end{eqnarray}%
where $p_{n}$ is the propobality that there are $n$ molecules in the lattice
site, $\gamma $ the molecular dissipation rate and $r_{a}=\left( T+\tau
\right) ^{-1}$ is the pumping rate of the fermionic atoms. Here, the
coefficient $\left\vert D_{2}\left( n\right) \right\vert ^{2}$ is defined as 
\[
\left\vert D_{2}\left( n\right) \right\vert ^{2}=1-\left\vert D_{1}\left(
n\right) \right\vert ^{2} 
\]%
where $\left\vert D_{1}\left( n\right) \right\vert ^{2}$ is defined in Eq. (%
\ref{p1}).

The steady state solution of Eq. (\ref{u2}) is 
\begin{equation}
p_{n}=p_{0}\prod_{l=1}^{n}\frac{\left\vert D_{2}\left( n-1\right)
\right\vert ^{2}N_{ex}}{l}  \label{u3}
\end{equation}%
where $p_{0}$ is the normalization constant and $N_{ex}=\left( r_{a}/\gamma
\right) $. Since the value $\left\vert D_{2}\left( n-1\right) \right\vert
^{2}$ is insensitive with the interaction time $\tau $, the steady molecular
distribution $p_{n}$ is also insensitive with $\tau $. A special case is
that $T$ is long enough (i.e. the detuning $\Delta $ is changed slowly
enough) so that the state $\left\vert e,n_{b}\right\rangle $ is adiabaticlly
changed into state $\left\vert g,n_{b}+1\right\rangle $ during each time
interval $\tau $. Therefore, we have $\left\vert D_{2}\left( n-1\right)
\right\vert ^{2}\approx 1$ which do not depend on both $\tau $ and $n$. In
this case the molecular distribution $p_{n}$ in Eq. (\ref{u3}) is
propotional to $N_{ex}^{n}/n!$. It is apparently that $p_{n}$ is just the
Possion distribution which is just the population distribution of a
molecular coherent state. The average molecular number $\left\langle
n\right\rangle $ is just $N_{ex}$ which can be controlled easily. Then we
can obtain a coherent output of molecule with the nonadiabatic transition
method. The coherence of the output molecular field can be described by the
line width $D$. This line width can be calculated with the standard method
as in the laser theory \cite{scully} and we have the result $D\approx
9\gamma /N_{ex}$. On the other hand, when the parameter $T$ is not very
large, it can be obtained with numerical simulation that we have $%
\left\langle n\right\rangle <N_{ex}$ and the Mandel $Q$ factor is larger
than $1$. Therefore, we can get superpossion distribution in this case.

In this paper we have shown that in the photoassociation process, we can
create the atom-molecule entanglement state with steady atomic and molecular
probability by varying the detunning of the Ramman laser beams. The
influence of the variation parameters of the detunning on the molecular
probability $P$ have also been investigated. As we have shown, if we want to
create a high probability $P$, the operation time ($4T$) should be long
enough. However, it can be seen from Fig. 4 that to obtain the maximal
entanglement state in Eq. (\ref{u1}), the operation time needed is much
shorter than the one needed to create a pure molecule state. We have also
shown that steady coherent output with Possion or superpossion molecular
distribution can be obtained by turning on and off the Raman laser beams for
many times. It is pointed out that, since in photoassociation process the
"Rabi frequency" $\chi $ is much larger than the molecular dissipation rate $%
\gamma $ \cite{weiping}, the condition that the time interval $\tau \lesssim
10^{2}\chi ^{-1}<<\gamma ^{-1}$ can be realized. Therefore, the scheme in
our paper can be used in the atom-molecule wave function engineering.

This work is supported by the CNSF (grant No.90203018), the knowledge
Innovation Program (KIP) of the Chinese Academy of Sciences and the National
Fundamental Research Program of China with No 001GB309310, the RGC grant of
Hong Kong (HKU7114/02P), the CRCG grant of HKU and the NSFC. We also
gratefully acknowledge the discussion with Prof. Weiping Zhang and the help
given by Y. Li.


\begin{references}
\bibitem[a]{email} Electronic address: suncp@itp.ac.cn

\bibitem[b]{www} Internet www site: http:// www.itp.ac.cn/\symbol{126}suncp

\bibitem{DK} u. N. Demkov and M. Kunike, Vestn. Leningr. Univ. Fis. Khim. 
{\bf 16}, 39 (1969).

\bibitem{models} K. A. Suominen and B. M. Garraway, Phys. Rev. A {\bf 45},
374 (1992).

\bibitem{p6} P. S. Julienne {\it et al}., Phys. Rev. A {\bf 58}, R797 (1998).

\bibitem{p7} D. J. Heinzen {\it et al}., Phys. Rev. Lett. {\bf 84}, 5029
(2000).

\bibitem{science} Roahn Wynar {\it et al}., Science {\bf 287}, 1016 (2000).

\bibitem{Mckenzie} C. Mckenzie {\it et al}., Phys. Rev. Lett. {\bf 88},
120403 (2002)

\bibitem{nature1} J. M. Gerton, {\it et al}., Nature (London) {\bf 417}, 529
(2002).

\bibitem{nature2} C. A. Regal, {\it et al}., Nature (London) {\bf 424}, 47
(2003).

\bibitem{rempe} S. Durr {\it et al}., Phys. Rev. Lett. {\bf 92}, 020406
(2004).

\bibitem{search} C. P. Search and P. Meystre, cond-mat/0404006

\bibitem{atom laser} M. O. Mewes {\it et al}., Phys. Rev. Lett. {\bf 78} 582
(1997).

\bibitem{lz} L. D. Landau, Phys. Z. Sowjetunion {\bf 2}, 46 (1932); C.
Zener, Proc. R. Soc. London Ser. A {\bf 137}, 696 (1932).

\bibitem{Kleppner} J. R. Rubbmark {\it et al}., Phys. Rev. A {\bf 23}, 3107
(1981).

\bibitem{weiping} C. P. Search {\it et al}., Phys. Rev. Lett. 91, 190401
(2003).

\bibitem{map} P. W. Anderson, Phys. Rev. {\bf 112}, 1900 (1958).

\bibitem{prove} It is easy to prove that $E_{e}, E_{a}$ and $E_{\pm}$
satisfy $dn_{b}/d(E_{e}+E_{a})>0$ and $E_+dE_{e}/dn_{b}>E_{e}dE_+/dn_{b}$
.Considering Eq. (\ref{t1}) and the inequality $y\tanh x<x\tanh y$ when $%
x>y>0$, it can be seen that these inequalities lead to the results $%
df/dn_{b}<0$ and $dg/dn_{b}<0$. Then we have the result $dP/dn_b>0$.

\bibitem{scully} M. O. Scully and M. S. Zubiary, Quantum Optics (Cambridge
University  Press, Cambridge, U.K., 1997).
\end{references}
\end{document}